\documentclass[doublecol,figures]{epl2}

\usepackage{amsmath}
\usepackage{graphicx}
\usepackage{amsmath}
\usepackage{amssymb}
\usepackage[british]{babel}
\usepackage{color}

\title{A solution to the subdiffusion-efficiency paradox: Inactive states
enhance reaction efficiency at subdiffusion conditions in living cells}
\shorttitle{A solution to the subdiffusion-efficiency paradox}

\author{Leila Esmaeili Sereshki\inst{1}
\and Michael A. Lomholt\inst{2}
\and Ralf Metzler\inst{3,4}\thanks{E-mail: \email{rmetzler@uni-potsdam.de}}}
\shortauthor{L. Esmaeili Sereshki \etal}

\institute{
\inst{1} Department of Physics, Technical University of Munich,
James-Franck Stra{\ss}e, 85747 Garching, Germany\\
\inst{2} MEMPHYS - Center for Biomembrane Physics,
Department of Physics, Chemistry and Pharmacy,
University of Southern Denmark, Campusvej 55,
5230 Odense M, Denmark\\
\inst{3} Institute for Physics and Astronomy, University of Potsdam,
D-14476 Potsdam-Golm, Germany\\
\inst{4} Department of Physics, Tampere University of Technology,
FI-33101 Tampere, Finland}

\date{\today}


\abstract{Macromolecular crowding in living biological cells effects
subdiffusion of larger biomolecules such as proteins and enzymes.
Mimicking this subdiffusion in terms of random walks on a critical
percolation cluster, we here present a case study of EcoRV restriction
enzymes involved in vital cellular defence. We show that due to its so far
elusive propensity to an inactive state the enzyme avoids non-specific
binding and remains well-distributed in the bulk cytoplasm of the cell.
Despite the reduced volume exploration capability of subdiffusion processes,
this mechanism guarantees a high efficiency of the enzyme.
By variation of the non-specific binding constant and the bond occupation
probability on the percolation network, we demonstrate that reduced non-specific
binding are beneficial for efficient subdiffusive enzyme activity even in
relatively small bacteria cells. Our results corroborate a more local picture
of cellular regulation.}

\begin{document}

\maketitle

\section{Introduction}
Diffusion-limited biochemical cellular reactions underlying signalling and
regulation processes have traditionally been investigated at dilute solvent
conditions \cite{bvh}. The relevance of this picture for diffusion control
in living cells has been challenged in view of \emph{macromolecular crowding},
the occupation of a considerable volume fraction $f$ of the cellular cytoplasm
by larger biopolymers \cite{minton}. Estimates for $f$ typically range from
35\% to 40\%. Bearing in mind that on a cubic lattice the site percolation
threshold is $f\approx31$\% and that of bond percolation $f\approx25$\%
\cite{havlin,ziff}, molecular crowding may indeed appear
severe.
Crowding effects changes in enzyme function and turnover, as well as protein
folding and aggregation \cite{mcguffee,zimmer}.

Larger biopolymers and tracers in living biological cells and artificially
crowded control environments perform subdiffusion of the form
\cite{golding,weber,seisenhuber,lene,garini,weiss1,weiss,pan,banks}
\begin{equation}
\label{msd}
\langle\mathbf{r}^2(t)\rangle\simeq t^{\alpha}\mbox{ with }0<\alpha<1,
\end{equation}
as observed experimentally for particles as small as 10 kD, with $\alpha$ in
the range of 0.40 to 0.90 \cite{pan,banks,weiss1,golding,weber},
in accord with
recent high-detail simulations \cite{mcguffee}. The observed
subdiffusion has
been measured to persist over tens to hundreds of seconds \cite{golding,weber}
and thus appears relevant to cellular processes such as gene regulation or
molecular
defence mechanisms. Subdiffusion leads to reduced global volume exploration,
to dynamic localisation at reactive interfaces \cite{zaid}, and may prevent
chromosomal mixing in eukaryotic nuclei \cite{garini}. It has been argued
that subdiffusion may in fact be beneficial for cellular processes, by
increasing the probability to find, and react with a \emph{nearby\/} target
\cite{golding,guigas}. In that sense subdiffusion would give rise to a more
local picture of diffusion-limited biochemical reactions in biological cells,
see below.

Here we study by simulations the reaction dynamics of EcoRV restriction enzymes
in \emph{Escherichia coli (E.coli)\/} bacteria. EcoRV in solution forms
homodimers of molecular weight 58 kD \cite{darcy}, and thus belongs to the range
of sizes for which subdiffusion (\ref{msd}) under crowding was reported
\cite{banks,weiss1,mcguffee}. Our results show that despite the subdiffusion
control
EcoRV's performance is surprisingly high. This provides a concrete solution to
the subdiffusion-efficiency paradox and supports current ideas that subdiffusion
does not contradict efficient molecular reactions in cells.

\section{EcoRV enzymes and our model approach}
The type II restriction endonuclease EcoRV binds non-specifically to
double-stranded DNA. Once it locates its specific six-base sequence
5'-GAT$|$ATC-3' it cuts the double-strand and renders it inactive.
This is an important mechanism in the cellular defence against alien DNA
stemming from, e.g., viruses attacking the cell \cite{jeltsch}. The cell's
native DNA is protected against EcoRV by methylation of the DNA at cytosine
or adenine \cite{pingoud}. Interestingly, EcoRV is found in two configurations
\cite{halford,winkler}: as seen by X-ray crystallography, the unbound protein
may switch between an inactive structure with a closed cleft and another, in
which the cleft is more open. In the open, active state EcoRV binds both
non-specifically to DNA and specifically to its cognate binding site.

\begin{figure}
\begin{center}
\includegraphics[width=8cm]{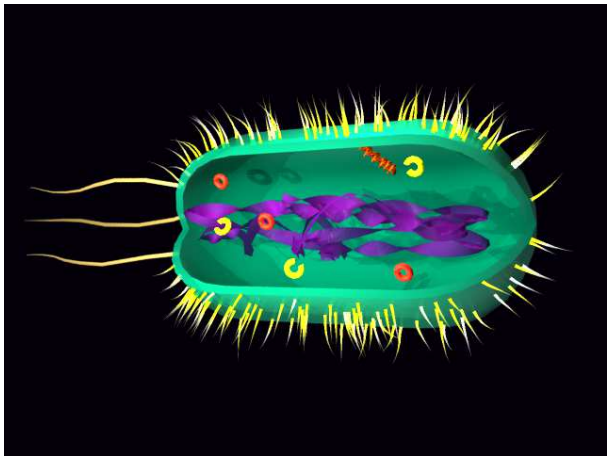}
\end{center}
\caption{Sketch of an \emph{E.coli\/} cell with native DNA (violet) concentrated
in the centre. EcoRV restriction enzymes occur in two isomers: inactive, with
closed cleft (red), and active with open cleft (yellow). Invading, foreign DNA
(red double-helix) is attacked by EcoRV and cut. The void intracellular space
shown here in reality is crowded by larger biopolymers.
\label{cell}}
\end{figure}

In Fig.~\ref{cell} we sketch an \emph{E.coli\/} cell and its native DNA, EcoRV
enzymes in active and inactive states being either attached non-specifically to
the native DNA or freely diffusing in the cellular cytoplasm. The apparent void
space in reality is a
highly crowded (`superdense' \cite{golding}) complex
liquid, in which the enzymes subdiffuse.
An invading DNA is being recognised an cleaved by active EcoRV enzymes.

Remarkably, the probability $x_{\mathrm{act}}$ to find the enzyme in the
open-cleft, active state at a given instant of time is as low as $\sim$1\%
\cite{gijs}. It is a priori puzzling why a vital defence mechanism should be
equipped with such a low activity. A physiologic rationale of the open/closed
isomerisation could be to reduce non-specific binding to the cell's native DNA.
Alien DNA invading the cell would thus immediately be surrounded by a higher
EcoRV concentration that, after switching to the active state, could attack this
DNA \cite{gijs}. Our results show, however, under the assumption of normal
diffusion in the cell the performance of EcoRV is only marginally better than
that of a 100\%-active mutant: normal diffusion on the length scales of an
\emph{E.coli\/} cell provides very efficient mixing, and the reduced activity
of EcoRV would not constitute an advantage. As we will show this
situation changes drastically under subdiffusion, and low activity in fact
becomes advantageous.

\begin{figure}
\begin{center}
\includegraphics[width=6.8cm]{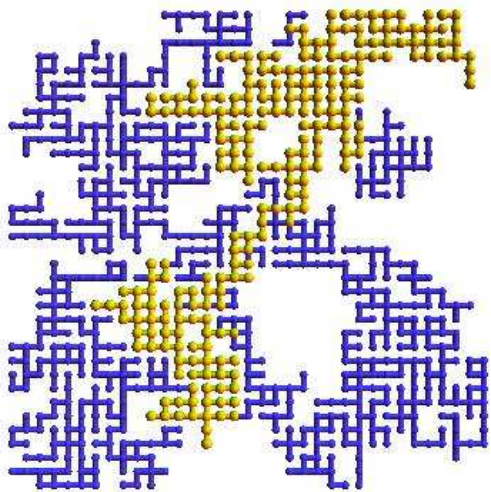}
\end{center}
\caption{Diffusion trace (yellow) on a bond percolation cluster.}
\label{perco}
\end{figure}

Following previous studies \cite{saxton1,benichou,langowski,REM} we model the
subdiffusion of EcoRV enzymes as random walks on a critical percolation cluster.
The use of a fractal medium is in line with observations that the crowded
cytoplasm may have a random fractal structure \cite{link1}. The lattice spacing
is chosen as the size of an EcoRV. A bond between lattice points is created with
occupation probability $p$. If $p=1$ the entire lattice is accessible and the
diffusion is normal. Reducing $p$ to the critical value $p_c=0.2488$ or slightly
above, the resulting cluster of permitted bonds is fractal with dimension $d_f
\approx2.58$ and the diffusion becomes anomalous with $\alpha\approx0.51$. From
extensive lattice simulations (see Appendix) we sample the times an enzyme needs
to locate its target, a specific sequence on an invading stretch of DNA randomly
positioned in the cellular cytoplasm (the volume not occupied by the native DNA).
The average target knockout time, equivalent to the mean first passage time
(MFPT) to hit the target in an active state, is studied as function of the bond
occupation probability $p$ and the non-specific binding constant $K^0_{\mathrm{
ns}}$ of active EcoRV to DNA.

Fig.~\ref{perco} depicts part of the spanning percolation cluster and part of
a random walk trace on this cluster. Due to the reduced connectivity in the
fractal cluster, the trajectory reflects the existence of holes existing on
all scales. The appearance of dead ends and bottlenecks in the scale-free
environment effects subdiffusion \cite{havlin}.

\section{Simulations results}
\begin{figure}
\begin{center}
\includegraphics[width=8.8cm]{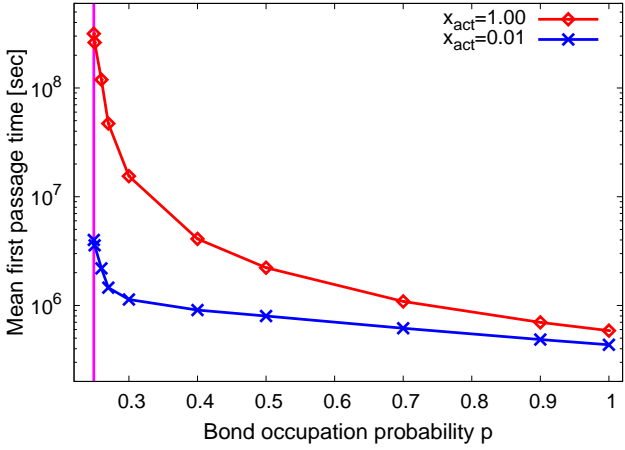}
\end{center}
\caption{Typical time for the restriction enzyme to locate the target in an
active state (MFPT) on a cubic lattice, as function of the bond occupation
probability. Close to criticality ($p_c=0.2488$ is marked by the vertical line),
subdiffusion emerges with anomalous diffusion exponent $\alpha=0.51$. The
non-specific binding constant is $K_{\mathrm{ns}}^0=10^7$ $[\mathrm{M}^{-1}
\mathrm{bp}^{-1}]$ \cite{supp}. Error bars are of the size of the symbols
or less.}
\label{fig:p}
\end{figure}

In Fig.~\ref{fig:p} we compare the MFPT for EcoRV (activity $x_{\rm act}=0.01$)
and mutant enzyme ($x_{\rm act}=1$) versus the bond occupation probability
$p$, ranging from full occupation ($p=1$, normal diffusion) down to the
percolation threshold $p=0.2488$ (subdiffusion with $\alpha=0.51$). For normal
diffusion ($p=1$) the MFPT is just a factor of two smaller for EcoRV, compared
to the fully active mutant. Approaching the percolation threshold the native
EcoRV increasingly outperforms the mutant, at criticality EcoRV's MFPT is
\emph{two orders of magnitude\/} shorter than that of the mutant.

On average, the concentration of EcoRV is approximately $1/x_{\rm act}=100$
times higher in
the cytoplasm outside the volume of the cell's native DNA than that of the
mutant enzyme. At criticality, it is time-costly to cover distances, and
thus EcoRV is 100 times more efficient than the fully active mutant enzymes.
The latter become trapped around the native DNA, to which they bind
non-specifically.
In contrast, under normal diffusion conditions ($p=1$) spatial separation
is hardly significant, and the lower concentration is compensated by the higher
activity of the mutant.

In absolute numbers, even under severe anomalous diffusion with $\alpha\approx
0.51$ EcoRV's MFPT is only a factor of ten higher than at normal diffusion.
That means that the low-activity property of EcoRV renders their efficiency
almost independent of the diffusion conditions, compared to the huge difference
observed for the mutant. The highly increased relative performance of the native
EcoRV is the central result of this study. It demonstrates that subdiffusion is
not prohibiting efficient molecular reactions. Moreover, our result provides a
novel rationale for EcoRV's elusive low-activity, that in this light appears as
a designed property. We note that the MFPT shown here is the result for an
individual EcoRV enzyme. Typically, a bacteria cell combines a fairly large
number of restriction enzymes of various families. This significantly reduces
the time scales indicated here, while preserving the characteristics of the
EcoRV superiority.
 
\begin{figure}
\begin{center}
\includegraphics[width=4.8cm]{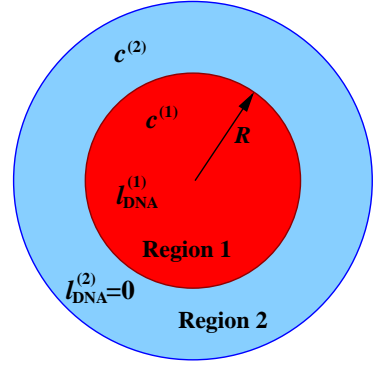}
\end{center}
\caption{Sketch of the cross section of an \emph{E.coli\/} model cell. Region 1
contains the cell's native DNA. In Region 2 (``cytoplasm''), foreign target DNA
are attacked by active restriction enzymes. The various symbols are explained in
the text.}
\label{sketch}
\end{figure}

\section{Discussion}
To obtain a better physical understanding of the r{\^o}le played by events of
EcoRV non-specific binding to the cell's native DNA, we study the dependence
of the MFPT on the non-specific binding constant $K^0_{\rm ns}$. Experimentally,
$K^0_{\rm ns}$ can be varied by changing the salt concentration of the solution.
For the cubic lattice ($p=1$) we obtain an analytical expression for the MFPT for
the geometry sketched in Fig.~\ref{sketch}. We distinguish Region 1 containing
the native DNA, and Region 2 representing the cytoplasm, in which the foreign
DNA enters and the EcoRV action occurs.

Let us first address the non-specific binding of EcoRV enzymes to the native
cellular DNA in Region 1, corresponding to a volume of length of $L$ and radius
$R$. Assuming rapid equilibrium with respect to enzyme binding and unbinding
from the DNA, we observe the following relation between the volume
concentrations of bound and unbound \emph{active\/} (ready-to-bind) enzymes,
\begin{equation}
\frac{c^{(1)}_{\rm act}}{c^{(1)}_{\rm bound}}=\frac{1}{K^0_{\rm ns}l^{(1)}_{\rm
DNA}}.
\end{equation}
In our notation $c^{(1)}$ is the overall volume concentration of
enzymes in Region 1, while $c^{(1)}_{\rm bound}$ and $c^{(1)}_{\rm bulk}$,
respectively, measure the volume concentrations of enzymes bound to the native
DNA and of unbound enzymes. The non-specific binding constant $K^0_{\rm ns}$ to
DNA refers to active (open-cleft) enzymes per DNA length, and is of dimension
$[K^0_{\rm ns}]=\mathrm{M}^{-1}\mathrm{bp}^{-1}$. Finally, $l^{(1)}_{\rm DNA}$
is the length of DNA per volume in Region 1. Of the unbound enzymes, a fraction
$x_{\rm act}$ is in the active (open-cleft) state, ready to bind to DNA. Thus,
the concentration of active unbound enzymes in Region 1 becomes $c^{(1)}_{\rm
act}=x_{\rm act}c^{(1)}_{\rm bulk}$, and one may introduce an overall binding
constant $K_{\rm ns}=x_{\rm act}K^0_{\rm ns}$:
\begin{equation}
\label{eq:equil2}
\frac{c^{(1)}_{\rm bulk}}{c^{(1)}_{\rm bound}}=\frac{1}{x_{\rm act}K^0_{\rm ns}
l^{(1)}_{\rm DNA}}=\frac{1}{K_{\rm ns}l^{(1)}_{\rm DNA}}.
\end{equation}
As the total enzyme concentration in Region 1 is $c^{(1)}=c^{(1)}_{\rm
bulk}+c^{(1)}_{\rm bound}$, and we have $c^{(1)}_{\rm bound}=c^{(1)}_{\rm
bulk}x_{\rm act}K^0_{\rm ns}l^{(1)}_{\rm DNA}$ we can write
$c^{(1)}_{\rm bulk}=c^{(1)}/[1+x_{\rm act}K^0_{\rm ns}l^{(1)}_{\rm DNA}]$.
In Region 1 the enzyme concentration in the continuum limit will be governed
by a diffusion equation of the form
\begin{equation}
\frac{\partial c^{(1)}}{\partial t}=D_{\mathrm{eff}}\nabla^2c^{(1)},
\end{equation}
where $D_{\rm eff}=D_{\rm 3d}/(1+x_{\rm act}K^0_{\rm ns}l^{(1)}_{\rm DNA})$
is an effective diffusion coefficient incorporating the assumption of rapid
equilibrium with respect to binding to DNA and switching between active and
dormant states. 1D diffusion along the DNA is assumed so slow that it can be
ignored in connection with the overall diffusion of the enzyme. Indeed the 1D
diffusion constant for EcoRV have been measured to be orders of
magnitude smaller than for 3D diffusion \cite{bonnet}.

In Region 2 we assume that 3D diffusion is fast allowing us to write a
conservation law for enzymes in the form of the difference between the flux
across the boundary with Region 1, and the amount of enzymes reacting with
the target per time,
\begin{equation}
\frac{d}{dt}V^{(2)}c^{(2)}=-A^{(1)}D_{\rm eff}\left.\frac{\partial c^{(1)}}{
\partial r}\right|_{r=R}-k_a c^{(2)}.
\end{equation}
Here $V^{(2)}$ is the volume of Region 2, $A^{(1)}=2\pi R L$ is the surface
area of Region 1, and $k_a$ is the rate constant for reaction with the target.
The $x_{\rm act}$ dependence of the rate 
$k_a$ is $k_a=x_{\rm act}k_a^0$, where $k_a^0$ is the rate constant
for the active state. Note that in our approach we assume that the switching
between active and dormant state is fast in comparison with the diffusion
across the regions, i.e., we may assume an equilibrium between these two states.
Finally, we take $c_{\rm bulk}$ to be continuous across the boundary between the
Regions 1 and 2, and that initially the system is at equilibrium with respect to
the reaction-free situation with $k_a=0$. From the above system of equations the
average search time yields in the form
\begin{equation}
\label{eq:MFPT}
T=\Big(1+x_{\rm act}K^0_{\rm ns}l^{(1)}_{\rm DNA}\Big)\left\{\frac{V^{(1)}}{
k_a^0 x_{\rm act}}(1+y)+\frac{R^2}{8D_{\rm 3d}}\frac{1}{1+y}\right\},
\end{equation}
where $y=V^{(2)}/\left[V^{(1)}(1+x_{\rm act}K^0_{\rm ns}l^{(1)}_{\rm DNA})
\right]$.

The simulations were carried out on a $100\times 100 \times 100$ cubic lattice
with native DNA occupying a $100\times 50 \times 50$ lattice in the middle. To
compare the present calculation with the simulations we assume a lattice spacing
of $a=10\,{\rm nm}$ and set $L=1\,\mu{\rm m}$, $V^{(2)}=1\,\mu{\rm m}^3$ and
thus $V^{(1)}=0.25\,\mu{\rm m}^3$. From this we obtain $R=\sqrt{V^{(1)}/(\pi L)}
\approx 0.28\,\mu{\rm m}$. Furthermore we choose the enzyme diffusivity $D_{\rm
3d}=3\mu\mathrm{m}^2/\mathrm{sec}$ [value obtained for lac repressor in vivo at
short times \cite{elf}] the DNA length per volume $l^{(1)}_{\rm DNA}=1.5\times
10^{-3}\mathrm{m}/V^{(1)}$, and $K^0_{\rm ns}=10^7\mathrm{M}^{-1}\mathrm{bp}^{
-1}$, such that $K_{\rm ns}=10^5\mathrm{M}^{-1}\mathrm{bp}^{-1}$ when $x_{\rm
act}=0.01$ \cite{jeltsch1} (except for plots where this parameter is varied)
with the base pair length
$\mathrm{bp}=0.35\,{\rm nm}$. For the target association rate constant we take
$k_a^0=(N a)^3/T_{\rm lattice}=a^3(1-R_{\rm 3d})/\tau_{\rm step}$. Here $T_{\rm
lattice}=N^3 \tau_{\rm step}/(1-R_{\rm 3d})$ is the average search time for a
random walker starting far from the target on a $N\times N\times N$ cubic
lattice and spending a time $\tau_{\rm step}$ per step to nearest neighbour
sites. $R_{\rm 3d}\approx 0.340537$ is the walker's return probability
to its origin \cite{hughes}. Matching $\tau_{\rm step}$ with the above diffusion
constant through the mean squared displacement of the walker we obtain $\tau_{\rm
step}=a^2/(6 D_{\rm 3d})\approx 5.6\,\mu{\rm s}$. The assumption of a one
lattice site target gives a target size of $a=10\,{\rm nm}$. We have taken this
target size to be lower than the in vitro effective sliding length \cite{gijs}
at optimal salt conditions, partly due to possible blocking on the DNA by
other DNA-binding proteins.

The above numbers yield $K^0_{\rm ns}l^{(1)}_{\rm DNA}\approx 3\times
10^5$, $y|_{x_{\rm act}=1}\approx 10^{-5}$, $y|_{x_{\rm act}=0.01}\approx
10^{-3}$, $V^{(1)}/k_a^0\approx 2\,{\rm sec}$, and $R^2/(8 D_{\rm 3d})\approx
0.003\,{\rm sec}$, and with these parameters we have to a good approximation
$T=K^0_{\rm ns}l^{(1)}_{\rm DNA} V^{(1)}/k_a^0$, regardless of whether $x_{\rm
act}=1$ or $x_{\rm act}=0.01$.

\begin{figure}
\begin{center}
\includegraphics[width=8.8cm]{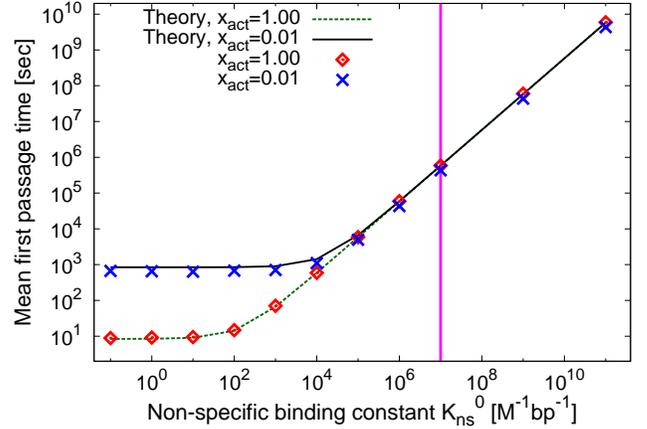}
\end{center}
\caption{MFPT on a normal lattice ($p=1$), as function of the non-specific
binding constant $K_{\mathrm{ns}}^0$. Simulations results are compared to the
theoretical result (\ref{eq:MFPT}). The vertical line marks the value $K_{
\mathrm{ns}}^0=10^7$ $[\mathrm{M}^{-1}\mathrm{bp}^{-1}]$ used in
Fig.~\ref{fig:p}.}
\label{fig:knserror}
\end{figure}

The simulations results for the case of normal diffusion are displayed in
Fig.~\ref{fig:knserror}. At small $K^0_{\rm ns}$ the mutant ($x_{\mathrm{
act}}=1$) clearly outperforms EcoRV, the gap in the MFPT corresponding to the
reduced activity ($x_{\mathrm{act}}=0.01$).
At increasing $K^0_{\rm ns}$ both EcoRV and mutant perform almost identically,
with a small advantage to EcoRV. In this regime almost all active enzymes are
bound to the cellular DNA, such that EcoRV has approximately a factor of
$1/x_{\mathrm{act}}=100$ higher bulk concentration. Concurrently, its
association rate constant with the target DNA in the cytoplasm is reduced by
the same factor. In this normally diffusive regime dominated by non-specific
binding, reduced activity of the restriction enzyme has no significant
advantage. The resulting MFPT behaviour according to Eq.~(\ref{eq:MFPT})
$T\approx K_{\mathrm{ns}}^0\ell_{\mathrm{DNA}}^{(1)}V^{(1)}/k_0^a$ in this
regime depends linearly on the non-specific binding constant. Indeed, this
behaviour is independent of $x_{\mathrm{act}}$. The agreement between the
theoretical model and the simulations results is excellent over the entire
range of $K^0_{\rm ns}$ (Fig.~\ref{fig:knserror}).

\begin{figure}
\begin{center}
\includegraphics[width=8.8cm]{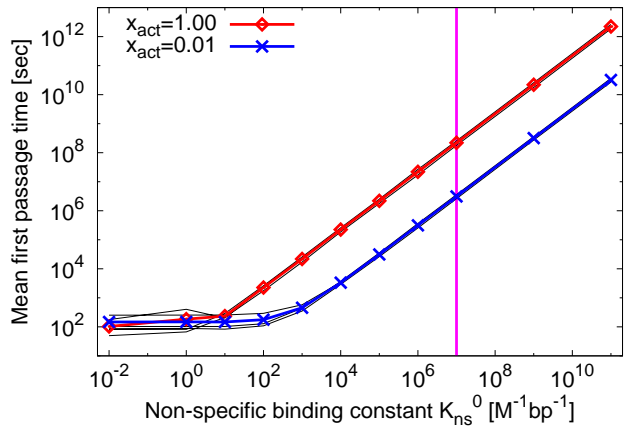}
\end{center}
\caption{MFPT on a percolation cluster close to criticality ($p_c=0.25$) versus
the binding constant $K_{\mathrm{ns}}^0$. The value $K_{\mathrm{ns}}^0
=10^7$ $\mathrm{M}^{-1}\mathrm{bp}^{-1}$ used in Fig.~\ref{fig:p} is marked
by the vertical line. Thick coloured lines: average over three
different percolation clusters. Thin black lines: results for the individual
clusters.}
\label{kns25}
\end{figure}

Fig.~\ref{kns25} shows the behaviour in the case of subdiffusion: almost over the
entire $K^0_{\mathrm{ns}}$ range, EcoRV significantly outperforms the mutant. At
sufficiently large $K^0_{\mathrm{ns}}$ values (above some $10^3$ $\mathrm{
M}^{-1}\mathrm{bp}^{-1}$) the value of the MFPT is approximately two orders of
magnitude smaller, i.e., the performance is improved by a factor close to the
value $1/x_{\mathrm{act}}$. This behaviour is thus dominated by the costly
subdiffusion from the site of non-specific binding to the target. At low $K^0_{
\mathrm{ns}}$ values both curves converge. Now, the MFPT is fully dominated
by anomalous diffusion to the target. Due to the compactness of the diffusion
on the fractal cluster the difference between EcoRV and the mutant becomes
marginal: upon an unsuccessful reaction attempt, EcoRV has a higher probability
to hit the target repeatedly before full escape, improving the efficiency.
In Fig.~\ref{kns25} the thick lines show the average of simulations over three
different critical percolation clusters, while the thin black lines depict the
result for each individual cluster. Apart from the low $K^0_{\mathrm{ns}}$
limit, the results are very robust to the shape of the individual cluster.
It would be interesting to derive analytically the MFPT dependence on $K^0_{
\mathrm{ns}}$. While the MFPT problem on a fractal has been solved recently
\cite{olivier}, it is not clear  how to apply this method in the present case,
due to the division of the support into two subdomains. Similarly, for the
related case of fractional Brownian motion \cite{chechkin} this remains an open
question.

\section{Conclusions} It is often argued that molecular processes in the cell
could not be subdiffusive, as this would compromise the overall fitness of
the cell due to the slowness of the response to external and internal
perturbations. Here we demonstrated a solution to this subdiffusion-efficiency
paradox: specific molecular design renders the efficiency of EcoRV enzymes
almost independent on the exact diffusion conditions. Even though EcoRV are
not always ready to bind, under subdiffusion conditions the low enzyme
activity represents a superior strategy.

Cellular subdiffusion may also be modelled by fractional Brownian motion (FBM)
or continuous time random walks (CTRW) \cite{pccp}. FBM shares many features
with diffusion on fractal structures, e.g., the compactness and ergodicity. The
essential observations for the MFPT found herein should therefore be similar
for the case of FBM. In contrast, for CTRW subdiffusion the high probability of
not moving in a given period of time will significantly enhance the advantage
of EcoRV over the mutant: while being trapped next to the target EcoRV would
have ample chance to convert to the active state and knock out the target.
It
will be interesting to compare our results to simulations on a dynamic
percolation cluster.

Subdiffusion-limited reactions generally increase the likelihood for biochemical
reactions to occur when the reactants are close-by \cite{golding,guigas}. Such
a more local picture of cellular biomolecular reactions in fact ties in with the
observed colocalisation of interacting genes \cite{mirny}. In higher cells the
similar locality is effected by internal compartmentalisation by membranes. It
will be interesting to obtain more detailed information from single particle
tracking experiments in living cells, in order to develop an integrated theory
for cellular signalling and regulation under crowding conditions in living
cells.

\acknowledgments

Financial support from the Deutsche Forschungsgemeinschaft, the Center for
Nanoscience, the Academy of Finland (FiDiPro scheme), and the Danish National
Research Foundation is gratefully acknowledged.

\section{Appendix}
Our simulations of the search process were carried out on a $100\times
100\times 100$ cubic lattice. We considered bond percolation, i.e., for each
pair of nearest neighbour sites a bond is constructed with a probability
$p$. The searching random walker is only allowed to walk between connected
sites (compare Fig.~\ref{perco}). The largest cluster of connected sites was
chosen and the remaining
sites discarded. Of these remaining sites those within the central
$100\times 50\times 50$-size part constitute Region 1 with the native DNA.
The number of such sites is denoted by $N^{(1)}$ and the volume of this region
is thus $V^{(1)}=a^3N^{(1)}$. The remaining sites outside this region
constitute the cytoplasm. The number of these sites is $N^{(2)}$, and the
corresponding volume becomes $V^{(2)}=a^3 N^{(2)}$. A single target site
is chosen randomly among the cytoplasm sites.

The initial position of the searching random walker is chosen by first deciding
whether it is bound or unbound. With probability $(V^{(1)}+V^{(2)})/[V^{(2)}+V^
{(1)}(1+x_{\rm act}K_{\rm ns}^0l_{\rm DNA}^{(1)})]$ it is chosen to be unbound.
In this case its initial position is chosen randomly among all $N^{(1)}+N^{(2
)}$ sites, and with probability $x_{\rm act}$ it is chosen to be in the active
state (otherwise it is in the inactive state). If the walker is initially
chosen to be bound, then it is placed randomly among the native DNA sites, and
its state is initially set as active.

The initial time is set to $t=0$, and the search is carried out according
to the following algorithm:

\begin{enumerate}

\item If the walker is in the inactive state or is situated in the cytoplasm,
a time $\tau_{\rm step}$ is added to the total time $t$. If the walker is
active and situated in the region with the native DNA, then a random time is
added to $t$. This latter random time is taken from an exponential distribution
with average $\tau_{\rm step}(1+K_{\rm ns}^0l_{\rm DNA}^{(1)})$.

\item One of the 6 directions possible on a cubic lattice is chosen at random.
If a bond exists to a neighbouring site in this direction, the walker moves to
this site, otherwise it stays put.

\item If the walker is in the active state, it switches with probability
1/5 to the inactive state. If instead the walker is in the inactive state,
it switches to the active state with probability $x_{\rm act}/5/(1-x_{\rm
act})$. In the case when $x_{\rm act}=1$, this switching is turned off and
the walker always stays in the active state.

\item If the walker is on the target site and in the active state, the target
is considered to be found, and the time $t$ is recorded as the search time.
Otherwise the iteration goes back to step 1.

\end{enumerate}

This procedure is repeated 5,000 times on the same percolation cluster, but with
a new target position and initial position of the searcher each time. The MFPT
is calculated as the average of the 5,000 recorded search times.


\begin{thebibliography}{99}

\bibitem{bvh} See, for instance, P. H. von Hippel and O. G. Berg,
%Facilitated target location in biological systems,
J. Biol. Chem. \textbf{264}, 675 (1989). For a recent example investigating
EcoRV, see Ref.~\cite{gijs} and M. A. Lomholt et al.,
%B. v. d. Broek, S.-M. J. Kalisch, G. J. L. Wuite, and R. Metzler, Facilitated
%diffusion with DNA coiling,
Proc. Natl. Acad. Sci. USA {\bf 106}, 8204 (2009).

\bibitem{minton} A. P. Minton,
%How can biochemical reactions within cells differ from those in test tubes?,
J. Cell Science \textbf{199}, 2863 (2006);
%\bibitem{huan}
H. Zhou, G. Rivas, and A. P. Minton,
%Macromolecular crowding and confinement: Biochemical, biophysical,
%and potential physiological consequences,
Ann. Rev. Biophys., \textbf{37}, 375 (2008);
%\bibitem{dix}
J. A. Dix and A.S. Verkman,
%Crowding effects on diffusion in solutions and cells,
Ann. Rev. Biophys., \textbf{37}, 247 (2008).

\bibitem{zimmer} S. B. Zimmerman and S. O. Trach,
%Estimation of macromolecule concentrations and excluded volume effects for
%the cytoplasm of escherichia coli,
J. Mol. Biol. \textbf{222}, 599 (1991).
%\bibitem{dong}
H. Dong, S. Qin, H.-X. Zhou,
%Effects of Macromolecular Crowding on Protein Conformational Changes,
PLoS Comp. Biol. \textbf{6}, e1000833 (2010).
%\bibitem{ellisminton}
R. J. Ellis and A. P. Minton,
%Cell biology - Join the crowd,
Nature, \textbf{425}, 27 (2003).
%\bibitem{ellis}
R. J. Ellis,
%Macromolecular crowding: obvious but underappreciated,
Trends Biochem. Sci. \textbf{26}, 597 (2001).
%\bibitem{zimmer2}
S. B. Zimmerman and A. P. Minton,
%Macromolecular crowding -
%biochemical, biophysical, and physiological consequences,
Annu. Rev. Biophys. Biomol. Struct. \textbf{22}, 27 (1993).

\bibitem{havlin} D. ben-Avraham and S. Havlin, Diffusion and Reactions in
Fractals and Disordered Systems (Cambridge University Press, Cambridge, UK,
2005).

\bibitem{ziff} C. D. Lorenz and R. M. Ziff, Phys. Rev. E \textbf{57}, 230
(1998).

\bibitem{mcguffee} S. R. McGuffee and A. H. Elcock,
%Diffusion, Crowding \&
%Protein Stability in a Dynamic Molecular Model of the Bacterial Cytoplasm,
PLoS Comput. Biol. \textbf{6}, e1000694 (2010).

%\bibitem{report} R. Metzler, and J. Klafter,
%The random walk's guide to anomalous diffusion: a fractional dynamics approach,
%Phys. Rep. \textbf{339}, 1 (2000).

\bibitem{seisenhuber} G. Seisenberger et al., Science \textbf{294}, 1929 (2001).
%G. Seisenberger, M. U. Ried, T. Endre{\ss}, H.
%B{\"u}ning, M. Hallek, and C. Br{\"a}uchle, Real-time single-molecule imaging
%of the infection pathway of an adeno-associated virus, Science \textbf{294},
%1929 (2001).

\bibitem{weiss1} M. Weiss, M. Elsner, F. Kartberg, and T. Nilsson,
%Anomalous subdiffusion is a measure for cytoplasmic crowding in living cells,
Biophys. J. \textbf{87}, 3518 (2004).

\bibitem{golding} I. Golding and E. C. Cox,
%Physical nature of bacterial cytoplasm,
Phys. Rev. Lett. \textbf{96}, 098102 (2006).

\bibitem{garini} I. Bronstein et al.,
%Y. Israel, E. Kepten, S. Mai, Y. Shav-Tal, E. Barkai, and Y. Garini, Transient
%Anomalous Diffusion of Telomeres in the Nucleus of Mammalian Cells,
Phys. Rev. Lett. \textbf{103},018102 (2009).

\bibitem{weber} S. C. Weber, A. J. Spakowitz, and J. A. Theriot,
%Subdiffusive motion of a polymer composed of subdiffusive monomers,
Phys. Rev. Lett. \textbf{104}, 238102 (2010).

\bibitem{lene} J.-H. Jeon et al.,
%V. Tejedor, S. Burov, E. Barkai, C. Selhuber,
%K. Berg-S{\o}rensen, L. Oddershede, and R. Metzler, In Vivo Anomalous
%Diffusion and Weak Ergodicity Breaking of Lipid Granules,
Phys. Rev. Lett. \textbf{106} 048103 (2011).

\bibitem{banks} D. Banks and C. Fradin,
%Anomalous diffusion of proteins due to molecular crowding,
Biophys. J. \textbf{89}, 2960 (2005).

\bibitem{pan} W. Pan et al.,
%L. Filobelo, N. D. Q. Pham, O. Galkin, V. V. Uzunova,
%and P. G. Vekilov, Viscoelasticity in Homogeneous Protein Solutions,
Phys. Rev. Lett. \textbf{102}, 058101 (2009).

\bibitem{weiss} J. Szymanski and M. Weiss,
%Elucidating the Origin of Anomalous Diffusion in Crowded Fluids,
Phys. Rev. Lett. \textbf{103}, 038102 (2009).

\bibitem{guigas} G. Guigas and M. Weiss,
%Sampling the cell with anomalous diffusion - The discovery of slowness,
Biophys. J. \textbf{94}, 90 (2008).

\bibitem{mirny} G. Kolesov et al.,
%Z. Wunderlich, O. N. Laikova, M. S. Gelfand, and L. A. Mirny, How gene order
%is influenced by the biophysics of transcription regulation,
Proc. Natl. Acad. Sci. USA \textbf{104}, 13948 (2007).

\bibitem{zaid} M. A. Lomholt, I. M. Zaid, and R. Metzler,
%Subdiffusion and weak ergodicity breaking in the presence of a reactive
%boundary,
Phys. Rev. Lett. \textbf{98}, 200603 (2007); I. M. Zaid, M. A. Lomholt,
and R. Metzler,
%How Subdiffusion Changes the Kinetics of Binding to a Surface,
Biophys. J. \textbf{97}, 710 (2009).

\bibitem{jeltsch} A. Jeltsch, C. Wenz, F. Stahl, and A. Pingoud,
%Linear diffusion of the restriction endonuclease EcoRV on DNA is essential for
%the in vivo function of the enzyme,
EMBO J. \textbf{15}, 5104 (1996).

\bibitem{pingoud} A. Pingoud, M. Fuxreiter, V. Pingoud, and W. Wende,
%Type II restriction endonucleases: structure and mechanism,
Cell. Mol. Life Sci, \textbf{62}, 685 (2005).

\bibitem{halford} S. G. Erskine, G. S. Baldwin, and S. E. Halford,
%Rapid reaction analysis of plasmid DNA cleavage by the EcoRV restriction
%endonuclease,
Biochem. \textbf{36}, 7567 (1997).

\bibitem{winkler} F. K. Winkler et al.,
%D. W. Banner, C. Oefner, D. Tsernoglou, and R. S. Brown, The crystal
%structure of EcoRV endonuclease and of its complexes with cognate and
%non-cognate DNA fragments,
EMBO J. \textbf{12}, 1781 (1993).

\bibitem{gijs} B. v. d. Broek et al.,
%M. A. Lomholt, S.-M. J. Kalisch, R. Metzler,
%and G. J. L. Wuite, How DNA coiling enhances target localization by proteins,
Proc. Natl. Acad. Sci. USA \textbf{105}, 41, 15738 (2008).

\bibitem{darcy} A. D'Arcy et al.,
%R. S. Brown, M. Zabeau, R. W. v. Resandt, and F. K.
%Winkler, Purification and crystallization of the EcoRV restriction endonuclease,
J. Biol. Chem. \textbf{260}, 1987 (1985).

\bibitem{link1} R. Cuthbertson, W. M. L. Holcombe, and R. Paton,Computation
in cellular and molecular biological systems (World Scientific, Singapore,
1995).

\bibitem{supp} See Supplementary Material.

\bibitem{saxton1} M. J. Saxton,
%Chemically limited reactions on a percolation cluster,
J. Chem. Phys. \textbf{116}, 203 (2002).

\bibitem{benichou} C. Loverdo, O. B{\'e}nichou, and R. Voituriez,
%Quantifying Hopping and Jumping in Facilitated Diffusion of DNA-Binding
%Proteins,
Phys. Rev. Lett. \textbf{102}, 188101 (2009).

\bibitem{langowski} C. C. Fritsch and J. Langowski, J. Chem. Phys.
\textbf{133}, 025101 (2010).

\bibitem{REM} Similar insights into the effect of crowding environments were
obtained from lattice random walks and off-lattice Brownian dynamics simulations
in J. D. Schmit, E. Kamber, and J. Kondev,
%Lattice Model of Diffusion Limited Bimolecular Chemical Reactions in
%Confined Environments,
Phys. Rev. Lett. \textbf{102}, 218302 (2009);
N. Dorsaz et al.,
%C. De Michele, F. Piazza, P. de los Rios, and
%G. Foffi, Diffusion Limited Reactions in Crowded Environments,
Phys. Rev. Lett. \textbf{105}, 120601 (2010).

\bibitem{bonnet} I. Bonnet et al., Nucleic Acids Res. \textbf{36}, 4118 (2008).

\bibitem{elf} J. Elf, G.-W. Li, X. S. Xie, Science {\bf 316}, 1191 (2007).

\bibitem{jeltsch1} A. Jeltsch and A. Pingoud, Biochemistry {\bf 37}, 2160 (1998).

\bibitem{hughes} B. D. Hughes, Random Walks and Random Environments, Vol. 1
(Oxford University Press, Oxford, UK, 1995).

\bibitem{olivier} S. Condamin, O. B{\'e}nichou, V. Tejedor,
R. Voituriez, and J. Klafter, Nature \textbf{450}, 77 (2007).

\bibitem{chechkin} J.-H. Jeon, A. V. Chechkin, and R. Metzler,
EPL \textbf{94}, 20008 (2011).

\bibitem{pccp} For a recent summary, see, e.g., S. Burov, J.-H. Jeon, R. Metzler,
and E. Barkai, Phys. Chem. Chem. Phys.  \textbf{13}, 1800 (2011).

\end{thebibliography}
\end{document}